\documentclass[12pt,aps,prd,preprint,tightenlines,superscriptaddress,
  showpacs,nofootinbib]{revtex4}
\newcommand{\PRE}[1]{{#1}} 

\usepackage{hyperref}      
\usepackage{bm}
\usepackage{epsfig}
\usepackage{color}
\usepackage{comment}
\usepackage{ulem}

\newcommand{\beqa}{\begin{eqnarray}}
\newcommand{\eeqa}{\end{eqnarray}}
\newcommand{\beq}{\begin{equation}}
\newcommand{\eeq}{\end{equation}}
\newcommand{\bay}{\begin{array}}
\newcommand{\eay}{\end{array}}

\newcommand{\ben}{\begin{enumerate}}
\newcommand{\een}{\end{enumerate}}
\newcommand{\bit}{\begin{itemize}}
\newcommand{\eit}{\end{itemize}}

\def\lt{\left}
\def\rt{\right}

\newcommand{\ifb}{\text{fb}^{-1}}

\newcommand{\gev}{\text{GeV}}
\newcommand{\tev}{\text{TeV}}

\newcommand{\s}{\text{s}}

\newcommand{\Mpc}{\text{Mpc}}

\newcommand{\eqref}[1]{Eq.~(\ref{#1})}
\newcommand{\eqsref}[2]{Eqs.~(\ref{#1}) and (\ref{#2})}

\newcommand{\secref}[1]{Sec.~\ref{sec:#1}}

\newcommand{\figref}[1]{Fig.~\ref{fig:#1}}
\newcommand{\figsref}[2]{Figs.~\ref{fig:#1} and \ref{fig:#2}}

\newcommand{\vev}[1]{\langle #1\rangle}

\newcommand{\mx}{m_\chi}
\newcommand{\mg}{m_{\tilde{G}}}
\newcommand{\mpl}{M_*}

\newcommand{\sgnmu}{{\text{sign}}(\mu)}
\newcommand{\thCP}{\theta_{\text{CP}}}

\renewcommand{\em}{\it}

\long\def\symbolfootnote[#1]#2{\begingroup%
\def\thefootnote{\fnsymbol{footnote}}\footnote[#1]{#2}\endgroup} 

\begin{document}

\preprint{UCI-TR-2012-06}

\title{ \PRE{\vspace*{1.5in}} 
Confluence of Constraints in Gauge Mediation: \\
The 125 GeV Higgs Boson and Goldilocks Cosmology 
\PRE{\vspace*{0.3in}} 
}

\author{Jonathan L.~Feng}
\affiliation{Department of Physics and Astronomy, University of
California, Irvine, CA 92697, USA
\PRE{\vspace*{.1in}}
}

\author{Ze'ev Surujon}
\affiliation{Department of Physics and Astronomy, University of
California, Irvine, CA 92697, USA
\PRE{\vspace*{.1in}}
}

\author{Hai-Bo Yu\PRE{\vspace*{.3in}}}
\affiliation{Michigan Center for Theoretical Physics, Department of
  Physics, University of Michigan, Ann Arbor, MI 48109, USA
\PRE{\vspace*{.4in}}
}

\date{May 2012}

\begin{abstract}
\PRE{\vspace*{.3in}} Recent indications of a 125 GeV Higgs boson are
challenging for gauge-mediated supersymmetry breaking (GMSB), since
radiative contributions to the Higgs boson mass are not enhanced by
significant stop mixing. This challenge should not be considered in
isolation, however, as GMSB also generically suffers from two other
problems: unsuppressed electric dipole moments and the absence of an
attractive dark matter candidate.  We show that all of these problems
may be simultaneously solved by considering heavy superpartners,
without extra fields or modified cosmology.  Multi-TeV sfermions
suppress the EDMs and raise the Higgs mass, and the dark matter
problem is solved by Goldilocks cosmology, in which TeV neutralinos
decay to GeV gravitinos that are simultaneously light enough to solve
the flavor problem and heavy enough to be all of dark matter.  The
implications for collider searches and direct and indirect dark matter
detection are sobering, but EDMs are expected near their current
bounds, and the resulting non-thermal gravitino dark matter is
necessarily warm, with testable cosmological implications.
\end{abstract}



\maketitle

\section{Introduction}

Recent results from the Large Hadron Collider (LHC) show intriguing
hints of what might be interpreted as a Higgs boson.  After having
analyzed more than $4~\ifb$ of integrated luminosity at 7 TeV, the
ATLAS and CMS Collaborations report excesses of diphoton events with
invariant mass around 125 GeV and local significances near
$3\sigma$~\cite{ATLAS:2012ae,Chatrchyan:2012tx}.  Further support for
this interpretation comes from exclusion ranges.  The combined ATLAS
and CMS data constrain the mass of a standard model (SM) Higgs boson
to be within three possible ranges, namely, $117.5-118.5~\gev$,
$122.5-127.5~\gev$, and above 543 GeV.  These results have profound
implications for physics beyond the SM.  In this study, we consider
the implications of these results for supersymmetry and, in
particular, supersymmetric models with gauge-mediated supersymmetry
breaking (GMSB)~\cite{Dine:1981za,Dimopoulos:1981au,Nappi:1982hm,%
  AlvarezGaume:1981wy,Dine:1994vc,Dine:1995ag}.

In supersymmetric theories, the Higgs boson's mass is generically low,
since its quartic coupling is determined by the electroweak gauge
couplings.  Radiative corrections may lift the Higgs boson's mass, but
in the minimal supersymmetric SM (MSSM), a Higgs boson mass near 125
GeV requires either large trilinear scalar couplings, leading to large
left-right stop mixing, or very large stop masses.  Without additional
structure, naturalness would seem to disfavor heavy stops, since they
imply significant fine tuning.  At the same time, heavy superpartners,
at least in the first and second generations, generically relax other
longstanding problems in supersymmetry, namely, those of unwanted
flavor and CP violation.  In fact, with generic flavor structures and
phases, bounds on flavor and CP violation require superpartner masses
to be much higher than even the masses preferred by the Higgs mass.  A
realistic and compelling supersymmetric model, then, should not only
accommodate a 125 GeV Higgs boson, but also address these
supersymmetric flavor and CP problems.

In GMSB models, the superpartner masses are generated by flavor-blind
gauge interactions, thereby solving the supersymmetric flavor problem
elegantly.  Such models are therefore highly motivated in ways that
generic supersymmetric theories, and particularly those with
gravity-mediated supersymmetry breaking, are not.  The recent Higgs
boson results present an interesting challenge for GMSB, however. In
GMSB, trilinear soft couplings typically vanish at the messenger
scale.  Although they are regenerated through renormalization group
(RG) evolution at the weak scale, their value is too small to play a
significant role in lifting the Higgs mass, requiring multi-TeV stops.
The recent Higgs results have therefore motivated many new GMSB
studies, which typically propose non-minimal field content to resolve
this tension~\cite{Meade:2008wd,Buican:2008ws,Evans:2011bea,%
  Draper:2011aa,Evans:2012hg,Kang:2012ra,Ajaib:2012vc,Ibe:2012dd,Shao:2012je}.

The Higgs mass constraints should not be considered in isolation,
however, as GMSB has other significant and longstanding challenges.
First, although GMSB elegantly suppresses flavor violation, it does
not generically suppress the CP-violating electric dipole moments
(EDMs) of the electron and neutron.  These EDM constraints are
stringent: for ${\cal O}(0.1)$ CP-violating phases, and using the
underlying GMSB relations to relate first and third generation
superpartner masses, current bounds on EDMs require the stop mass to
be larger than $\sim 3-10~\tev$, depending on $\tan\beta$. Second,
typical GMSB models have no viable dark matter candidates, as all SM
superpartners decay to gravitinos, and thermally-produced
gravitinos~\cite{Pagels:1981ke} are inconsistent with standard big
bang cosmology.

In this work, we note that all of these problems are simultaneously
solved by having heavy superpartners.  In fact, taking minimal GMSB as
a simple example, the constraints all point to the same region of
parameter space.  As indicated above and detailed below, the Higgs
boson mass and EDMs point to the same range of multi-TeV superpartner
masses.  Remarkably, in this region of parameter space, the dark
matter problem is also solved by Goldilocks
cosmology~\cite{Feng:2008zza}, a superWIMP scenario, in which
TeV-scale neutralinos freeze out with very large densities, but then
decay to GeV gravitinos that are simultaneously light enough to solve
the flavor problem and heavy enough to be all of dark matter.  This
scenario is subject to many additional astrophysical
constraints~\cite{Feng:2003xh,Feng:2003uy,Kaplinghat:2005sy,%
  Cembranos:2005us}: the resulting gravitino dark matter should have
the correct relic density and be sufficiently cold, and
electromagnetic and hadronic energy produced in the decays should not
destroy the successes of big bang nucleosynthesis (BBN).  As we will
see, even more remarkably, all of these constraints are also satisfied
in the same region in GMSB parameter space preferred by the Higgs and
EDM constraints, without the need to modify standard big bang
cosmology.  

In \secref{gmsb}, we discuss the implications of recent Higgs data for
supersymmetry and minimal GMSB in particular.  The generic CP problem
of GMSB and the implications of bounds on EDMs are discussed in
\secref{edm}.  In \secref{cosmo} we consider dark matter in GMSB, and
discuss constraints from relic density, small scale structure, and
BBN.  As we will see, although the scenario we propose passes all
constraints, for several observables, the favored region of parameter
space is not far from current bounds.  There are therefore several
avenues where future sensitivities will be able to test these ideas,
and we discuss these and conclude in \secref{conc}.

\section{The Higgs Boson Mass and Minimal GMSB}
\label{sec:gmsb}

As mentioned in the introduction, in the MSSM the Higgs boson is
generically light, since the quartic coupling in the scalar potential
arises from $D$-terms and is therefore determined by the electroweak
gauge couplings.  Indeed, the tree level value,
\begin{equation}
   m_h^2({\text{tree}})=M_Z^2 \cos^2 2 \beta \ ,
   \label{treehiggs}
\end{equation}
cannot exceed the $Z$ boson mass.  This feature is retained even when
supersymmetry is softly broken, since quartic couplings are
dimensionless.  Radiative corrections, however, may quite generally
lift the value of $m_h^2$ by as much as 100\%.  The 1-loop correction
to \eqref{treehiggs} is given by
\begin{eqnarray}
\label{higgs}
   \Delta m_h^2({\text{1-loop}})=
   \frac{3 m_t^4}{2 \pi^2 v^2}
   \left[\log\left(\frac{M^2_S}{m_t^2}\right)
   +\frac{X^2_t}{M^2_S}\left(1-\frac{X^2_t}{12M^2_S}\right)\right] \ ,
\end{eqnarray}
where $v \simeq 246~\gev$, $M_S \equiv \sqrt{m_{\tilde{t}_1}
  m_{\tilde{t}_2}}$, and $X_t \equiv A_t - \mu \cot\beta$
characterizes the stop left-right mixing.  In addition, there are
higher-loop contributions that are known to be sizable.  Throughout
the paper we use \texttt{SOFTSUSY 3.2.4}~\cite{Allanach:2001kg} to
calculate the superpartner spectrum and Higgs boson mass, including
2-loop corrections and renormalization group (RG) evolution.  

Although \eqsref{treehiggs}{higgs} are modified significantly by
higher-loop corrections, they reveal a few interesting features of the
Higgs sector in the generic MSSM.  First, increasing $\tan\beta$
increases the tree-level Higgs mass, an effect that saturates for
$\tan\beta \sim 20$.  It is also evident that the Higgs boson mass may
be greatly increased either by large stop mixing ($X_t \sim M_S$) or
by heavy stops ($M_S\gg m_t$).  The large stop mixing scenario has
been investigated in many papers recently.  However, as we have
reviewed in the introduction, there is ample motivation from
considerations of flavor and CP violation to consider heavy
superpartners.

As a particularly simple example, we consider minimal GMSB.  In
minimal GMSB, the low energy spectrum is completely determined by the
five parameters
\begin{equation}
M_m , \ \Lambda \equiv \frac{F}{M_m} , \ 
\tan\beta \equiv \frac{\vev{H^0_u}}{\vev{H^0_d}} , \
N_5, \ \text{and} \ \, \sgnmu \ ,
\end{equation}
where the first is the messenger mass, $\Lambda$ (multiplied by a loop
factor $\sim \alpha/4\pi$) parameterizes the superpartner mass scale,
and the last two are discrete parameters that denote the equivalent
number of $5+\bar 5$ messengers and the sign of the Higgsino mass
parameter.  Regarding the Higgs potential, the soft scalar mass
parameters $m^2_{H_u}$ and $m^2_{H_d}$ are completely determined by
GMSB (and are essentially the same as the slepton doublet masses).
The Higgsino mass parameter $\mu$ and the soft bilinear parameter
$B_\mu$ are more problematic to generate.  Here, we follow the
traditional approach: we assume $\mu$ and $B_\mu$ are generated such
that $v\simeq 246~\gev$, and we trade them for $\tan\beta$ and $v$.
Assuming $\mu$ is real, the resulting free parameters are $\tan\beta$
and $\sgnmu$.  

Because the above parameters are flavor blind, the resulting low
energy physics is minimally flavor violating, and therefore safe from
flavor problems, as long as contributions from gravity-mediated
supersymmetry breaking are small compared to the superpartner mass
scale.  The gravity-mediated contributions are of the order of the
gravitino mass
\begin{equation}
   \mg=\frac{F}{\sqrt{3}\mpl}
   =\frac{M_m\Lambda}{\sqrt{3}\mpl} \ ,
\end{equation}
where $\mpl \simeq 2.4 \times 10^{18}~\gev$ is the reduced Planck
mass, and so the latter condition may be taken to be $\mg \ll (\alpha
/ 4 \pi) \Lambda$.  {}From this it follows that the gravitino is the
lightest supersymmetric particle (LSP).  As for CP violation, the
situation is less predictive, as we discuss in \secref{edm}.

One distinctive feature of minimal GMSB is that the $A$-terms vanish
at the messenger scale. Although they acquire a renormalization
contribution proportional to the gaugino mass at low energy scales,
their values are typically small, which, in light of the recent Higgs
signals, implies large stop masses.

To see the parameters required to generate a 125 GeV Higgs mass in
minimal GMSB models, we present results for minimal GMSB in
\figref{higgsedm}.  For $\tan\beta = 10$, a Higgs mass in the range of
$122.5-127.5~\gev$ implies $\Lambda \sim 700 - 3000~\tev$ and stop
masses $M_S \sim 4.5 - 20~\tev$, with smaller and much larger values
required for other values of $\tan \beta \agt 5$. These results are
for $N_5=1$ and $\mu > 0$.  Choosing other values for $N_5$ and
$\sgnmu$ would lead to different values of $\Lambda$ and would induce
some changes in the details of the spectrum, but, of course, the same
values of $M_S$ would be required, and this would not change our
conclusions qualitatively.  The value of the top mass is taken to be
$m_t=173.2~\gev$, the most recent value from the
Tevatron~\cite{Brandt:2012ui}. The ATLAS Collaboration has recently
measured a central value of $m_t=174.5~\gev$ with a statistical error
similar to that of the Tevatron combination, but with a larger
systematic error~\cite{topmassatlas}.

\begin{figure}[tbp]
\centering
\includegraphics[width=0.495\textwidth]{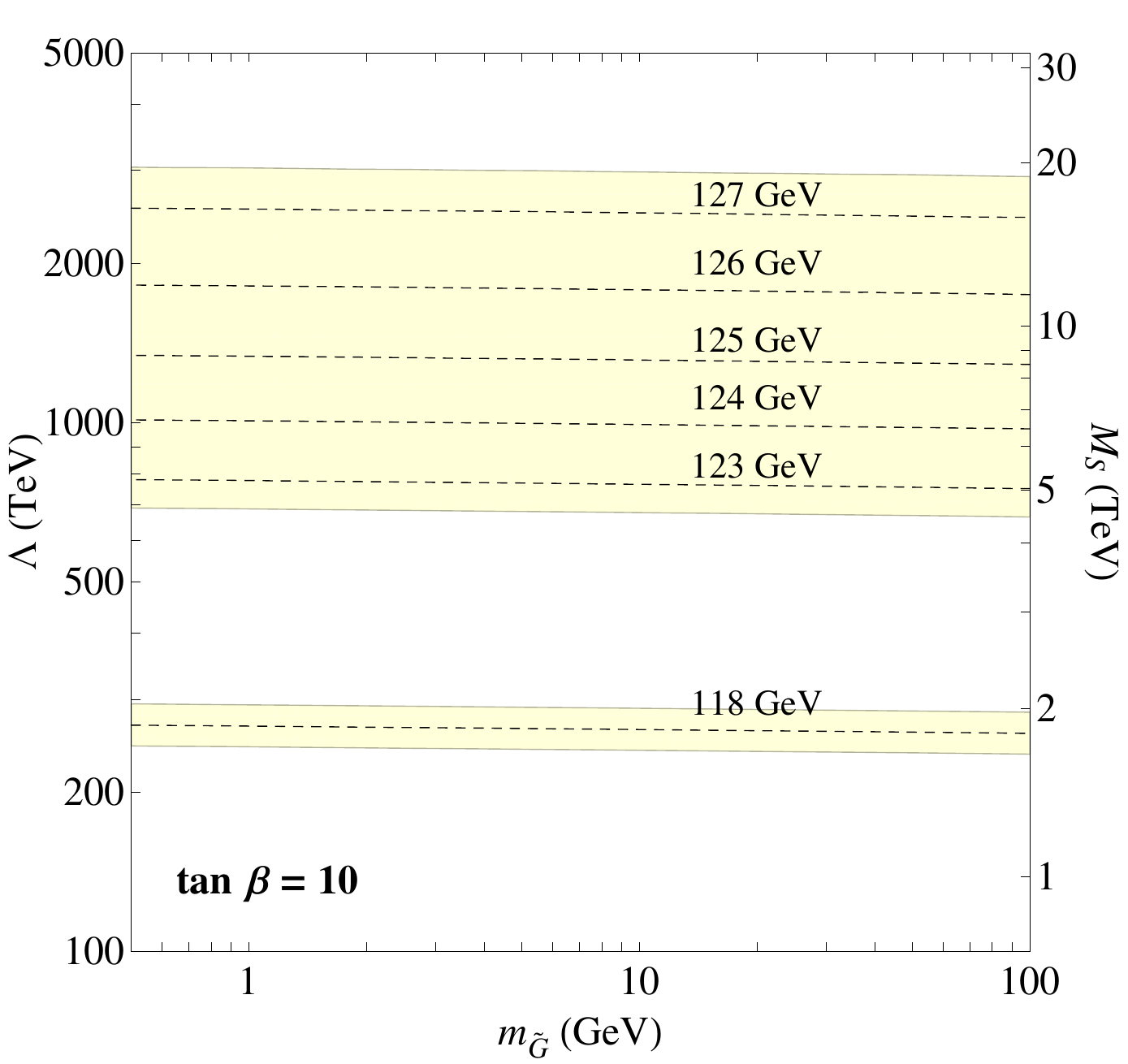}
\hfil
\includegraphics[width=0.495\textwidth]{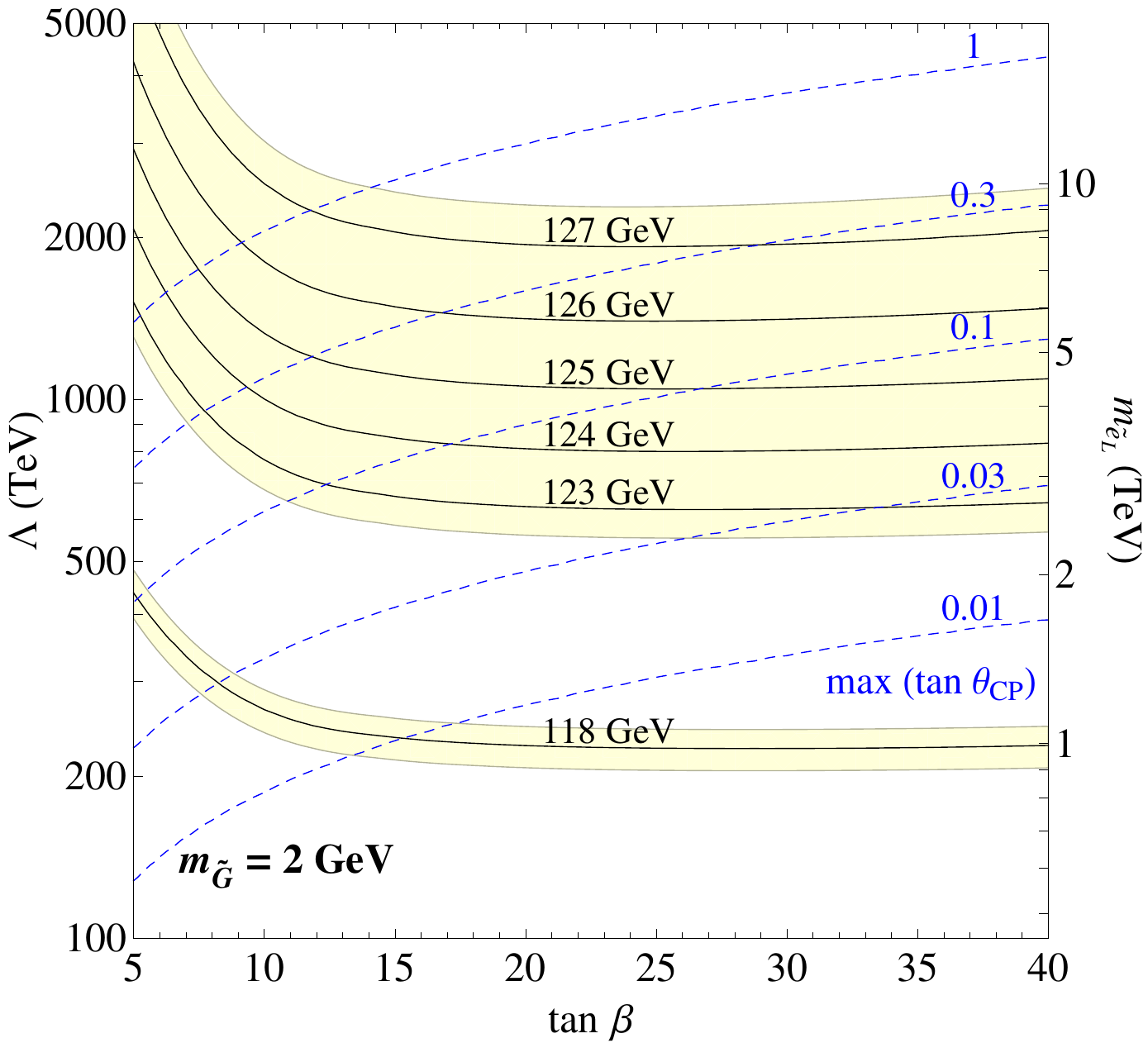}
\caption{\label{fig:higgsedm} {\em Left}: Contours of constant Higgs
  boson mass in minimal GMSB in the $(\mg, \Lambda)$ plane for
  $\tan\beta=10$, $N_5 = 1$, and $\mu > 0$.  The yellow shaded regions
  are the allowed ranges for the Higgs boson mass, given the recent
  exclusions from the LHC.  The stop mass parameter $M_S$ is largely
  determined by $\Lambda$ and insensitive to $\mg$ in the range
  plotted, and it is given on the right-hand axis.  {\em Right}:
  Dashed blue contours of constant maximal $\tan \thCP$ allowed by the
  upper bound on the electron EDM in the $(\tan\beta, \Lambda)$ plane
  for $\mg = 2~\gev$, $N_5 = 1$, and $\mu > 0$.  The preferred Higgs
  mass regions are as in the left panel.  The left-handed selectron
  mass $m_{\tilde{e}_L}$ is largely determined by $\Lambda$ and
  insensitive to $\tan\beta$, and it is given on the right-hand
  axis. }
\end{figure}

\section{Electric Dipole Moments}
\label{sec:edm}

In this section, we show that multi-TeV sfermion masses are also
motivated by constraints from EDMs. Although flavor violation is
highly suppressed in GMSB models, CP violation is not.  In the absence
of some additional mechanism to suppress CP
violation~\cite{Dine:1996xk,Moroi:1998km}, the gaugino masses $M_a$,
$A$-terms, and the $\mu$ and $B_\mu$ parameters can all have
CP-violating phases. In minimal GMSB, where the gaugino masses have
the same phase and $A$-terms vanish, the physical CP-violating phase
can be parameterized as
\begin{equation}
   \thCP\equiv \text{Arg}\lt(\frac{\mu M_a}{B_\mu}\rt) \ . 
\end{equation}

The EDMs of the electron and neutron are generated by penguin diagrams
with gauginos, Higgsinos and sfemions in the loop.  The dominant
diagram involves Wino-Higgsino mixing and leads to the EDM
contribution~\cite{Feng:2001sq}
\begin{equation}
d_f = \frac{1}{2} e \, m_f \, g_2^2 \, |M_2\,\mu| \, \tan \beta \,
\sin \thCP \,
K_C \lt( m_{\tilde{f}_L}^2, |\mu|^2, |M_2|^2\rt) \ ,
\end{equation}
where $K_C$ is a kinematic function defined in
Ref.~\cite{Moroi:1995yh}.  Note the factor of $\tan\beta$, which
arises from the down-type mass insertion required by the chiral
structure of the EDM operator.

The current upper bounds on electron and neutron EDMs
are~\cite{Hudson:2011zz,Baker:2006ts}
\begin{equation}
d_e<1.05\times10^{-27}~e~\text{cm} \quad 
\text{and} \quad 
d_n < 2.9 \times 10^{-26}~e~\text{cm} \ .
\end{equation}
Since the bound on $d_e$ is stronger than that on $d_n$, and the
theoretical value of $d_n$ is suppressed by heavier squark masses in
minimal GMSB, we will focus on the electron EDM.  Instead of
calculating the EDM diagram directly, we take advantage of the
similarity between the EDM and magnetic dipole moment operators.  We
first use \texttt{micrOMEGAs
  2.4.5}~\cite{Belanger:2001fz,Belanger:2004yn}, suitably modified to
include GMSB, to extract the anomalous magnetic moment of the muon
$a_\mu$.  The electron EDM is then given by
\begin{equation}
d_e = e\, \frac{m_e}{2m^2_\mu}\, a_\mu\tan\thCP \ ,
\end{equation}
where $e\equiv\sqrt{4\pi\alpha}$, $m_e$ is the electron's mass, and
$m_\mu$ is the muon's mass.  The maximal values of $\tan \thCP$
allowed by the electron EDM bound are shown in \figref{higgsedm} in
the $(\tan\beta, \Lambda)$ plane.  The dependence on $\tan \beta$ and
$\Lambda$ imply that regions with low $\tan\beta$ and large $\Lambda$
(and more generally, large $M_S$) are preferred. It is interesting to
note that the parameter space that gives rise to a 125 GeV Higgs boson
mass coincides with the region preferred by EDM considerations,
although these two constraints originate from completely different
sources.

\section{Dark Matter}
\label{sec:cosmo}

The standard dark matter candidate in supersymmetry is the neutralino,
which freezes out with the desired relic density naturally.  This
coincidence, the WIMP miracle, is not found in gauge mediation,
because the LSP is the gravitino.  The original thermally-produced keV
gravitino dark matter possibility is also no longer consistent with
the standard cosmological picture, as it is excluded by overclosure
and small-scale structure constraints~\cite{Feng:2010ij}.

GMSB may, however, give rise a viable dark matter scenario if
gravitino dark matter is produced non-thermally in neutralino
decays~\cite{Feng:2008zza}.\footnote{For related work with GeV-scale
  dark matter produced in late decays of TeV-scale particles, see
  Refs.~\cite{Kitano:2004sv,Kitano:2005ge,Ibe:2006rc,Ibe:2007km}.}  In
this scenario, the neutralino first freezes out with a large
abundance, and then decays to the gravitino. The resulting gravitino
inherits the neutralino's number density, but its energy density is
given by
$\Omega_{\tilde{G}}h^2=(\mg/m_\chi)\Omega_{\chi}h^2$. Although $\mg$
must be much less than $m_{\chi}$ to preserve the flavor virtues of
GMSB, this scenario realizes the WIMP miracle as much as is possible
in a GMSB visible sector,\footnote{If dark matter arises from hidden
  sectors in GMSB, a related ``WIMPless miracle'' may produce the
  desired amount of dark
  matter~\cite{Feng:2008ya,Feng:2008mu,Feng:2009mn}.} in the sense
that the final dark matter density is brought to near its desired
value by the thermal freezeout of a WIMP.  In this section, we map out
the allowed parameter space for gravitino dark matter and consider
cosmological constraints.

In GMSB with $N_5=1$, the NLSP is a Bino-like neutralino throughout
parameter space. Because neutralinos are Majorana particles, their
annihilation to quarks and leptons is $P$-wave suppressed. For the
Bino-like neutralino, its annihilation to gauge and Higgs bosons is
also suppressed. Because of these effects, the neutralino density at
freezeout may easily reach very large values. To see this, we use
\texttt{MicrOMEGAs 2.4.5} to calculate the thermal relic density the
neutralinos would have had had they been stable. This calculation
includes all annihilation channels and effects from non-Bino
contributions.  In \figref{binorelic}, we show the freezeout density
of neutralinos $\Omega_{\chi} h^2$. As expected, the neutralino
freezeout density is much larger than the observed dark matter density
throughout the parameter space. For $\tan\beta=10 \ (30)$, the
neutralino mass is close to 2 TeV (1.5 TeV) and its density is
$\Omega_\chi h^2\sim 100\ (50)$ in the region preferred by the Higgs
mass constraints.

\begin{figure}[tbp]
\centering
\includegraphics[width=0.495\textwidth]{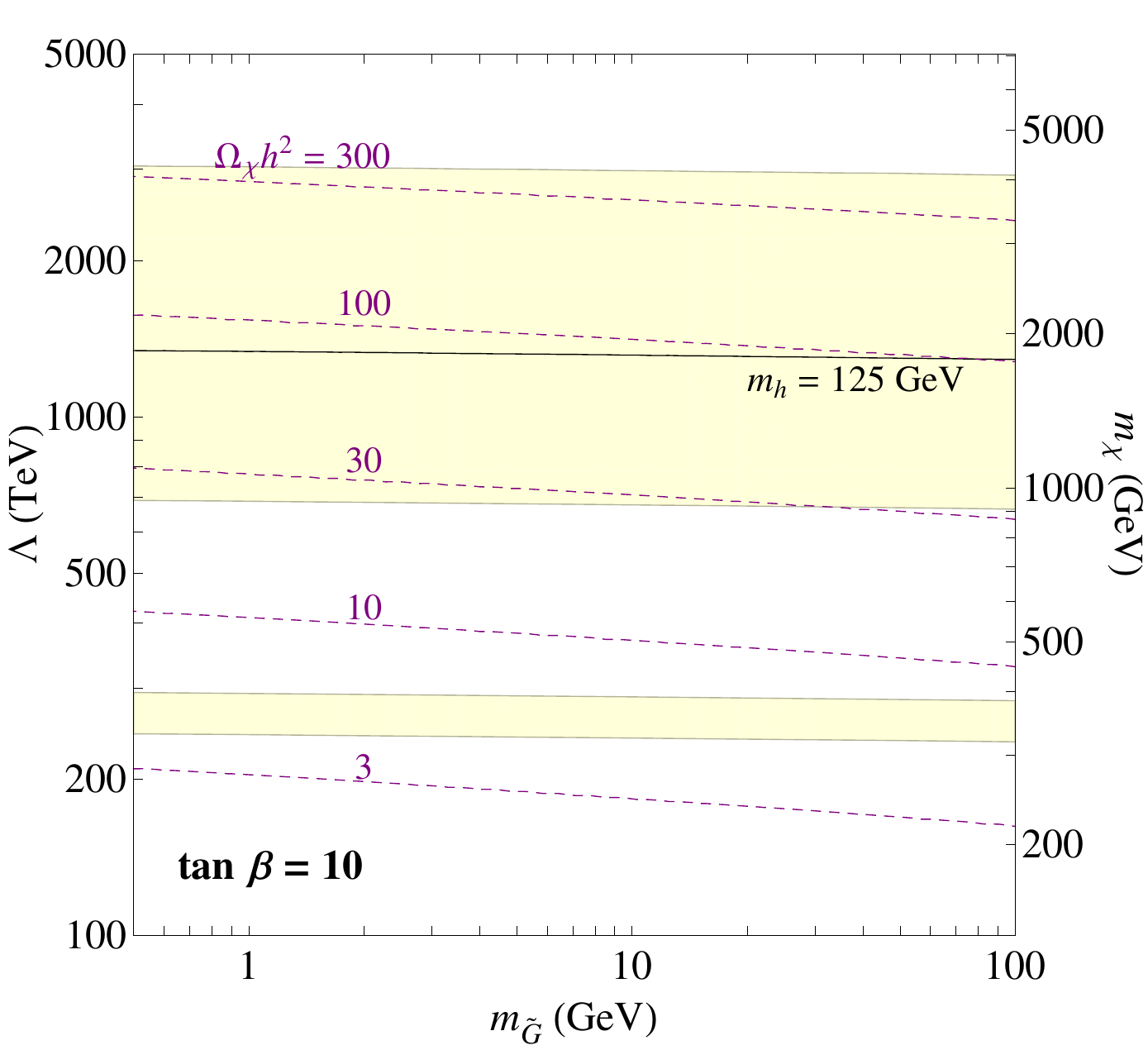}
\hfil
\includegraphics[width=0.495\textwidth]{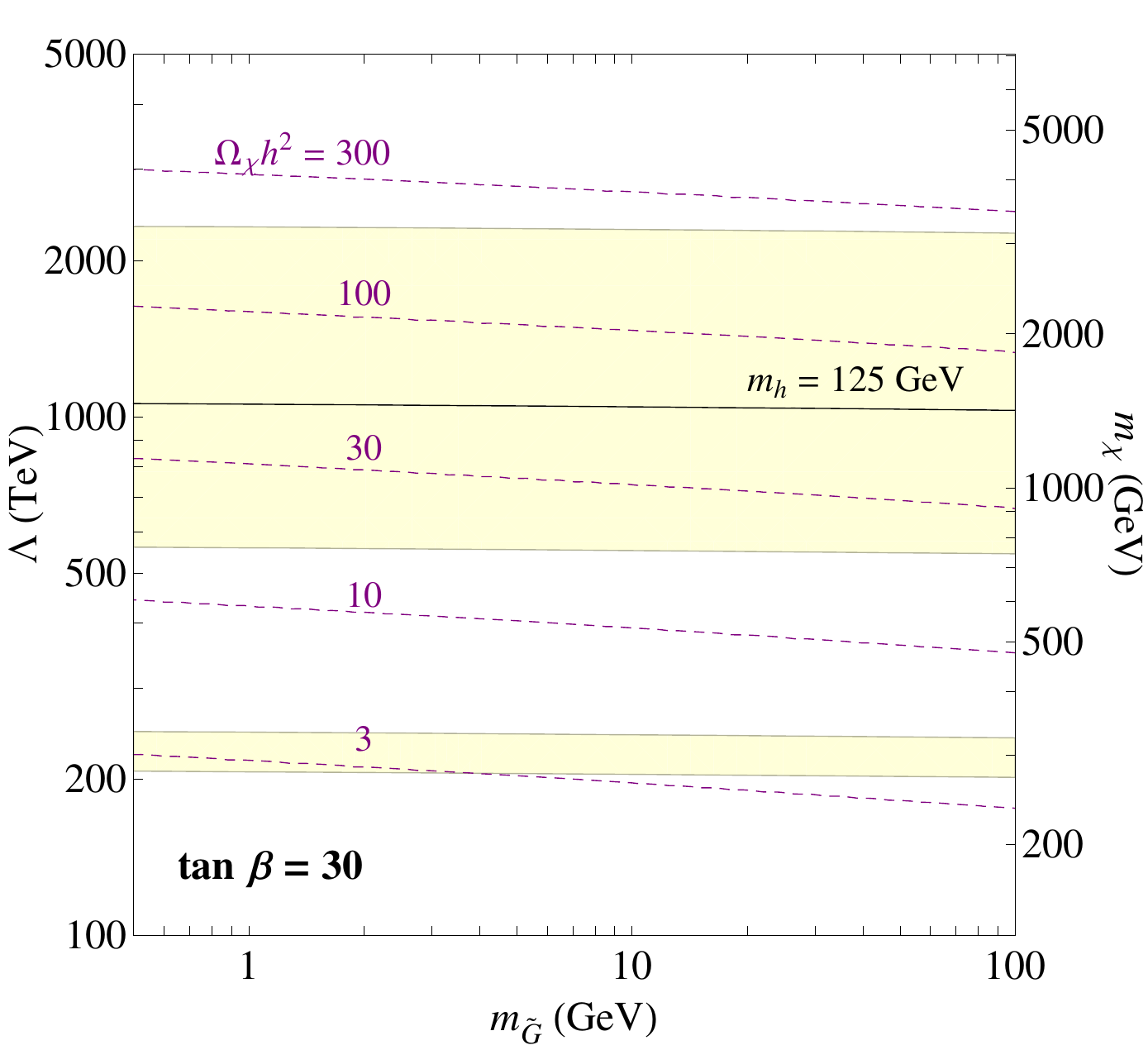}
\caption{\label{fig:binorelic} Contours of constant neutralino relic
  density (dashed, purple) for $\tan \beta = 10$ (left) and 30
  (right), $N_5 = 1$, and $\mu > 0$, computed with
  \texttt{MicrOMEGAs}.  The preferred Higgs mass regions from
  \figref{higgsedm} are also shown.  The neutralino mass $m_{\chi}$ is
  largely determined by $\Lambda$ and insensitive to $\mg$ in the
  range plotted, and it is given on the right-hand axis. }
\end{figure}

These freezeout densities are, however, not current relic densities,
as neutralinos in GMSB are unstable and decay to gravitinos. In
\figref{relic}, we show values of the gravitino relic density
$\Omega_{\tilde{G}}h^2$ along with the Higgs mass preferred regions
and other constraints discussed below. The region with viable
gravitino dark matter, where $\Omega_{\tilde{G}}h^2 = 0.112 \pm
0.006$, is a narrow band.  The slope of the band can be understood by
dimensional analysis. The freezeout density of neutralinos is
inversely proportional to the annihilation cross section, so
$\Omega_\chi h^2 \propto \vev{\sigma v}^{-1} \sim \tilde{m}^2$, where
$\tilde{m}$ is the superpartner mass scale. The gravitino relic
density is $\Omega_{\tilde{G}} h^2 = (\mg/\tilde{m})
\Omega_{\chi}h^2\sim\mg\tilde{m}$.  Roughly we have
$\tilde{m}\propto\Lambda$, so $\Omega_{\tilde{G}}h^2\sim\mg\Lambda$.
The gravitino masses that yield the correct relic density are in the
range $\mg \sim 1-10~\gev$. Such masses correspond to ``high-scale
GMSB,'' but are low enough to preserve the elegant flavor suppression
that motivates GMSB.

In the above discussion, we have assumed that the relic gravitino dark
matter is completely generated by neutralino decay after it freezes
out. As is well known, if the reheating temperature is high, inelastic
scattering processes can convert SM particles to gravitinos
efficiently~\cite{Weinberg:1982zq,Ellis:1984eq,Moroi:1993mb,Bolz:2000fu}.
The gravitino relic density produced through these processes during
reheating is approximately~\cite{Bolz:2000fu}
\begin{equation}
\Omega_{\tilde{G}} h^2 \approx 0.13 
\left(\frac{T_R}{10^6~{\gev}}\right)
\left(\frac{1~\gev}{\mg}\right)
\left(\frac{m_{\tilde g}}{7~\tev}\right)^2 \ ,
\end{equation}
where $T_R$ is the reheating temperature, and $m_{\tilde g}$ is the
running gluino mass. If the reheating temperature is significantly
less than $10^6~\gev$, the gravitino density produced by inelastic
scattering in the thermal bath is negligible. Of course, we require
also that $T_R$ is large enough that neutralinos are initially in
thermal equilibrium.  There is a large range of $T_R$, however, in
which both conditions are satisfied, gravitino dark matter is
dominantly from neutralino decays, and Goldilocks cosmology is
realized, thereby keeping the virtues of the WIMP miracle.

\begin{figure}[tbp]
\centering
\includegraphics[width=0.495\textwidth]{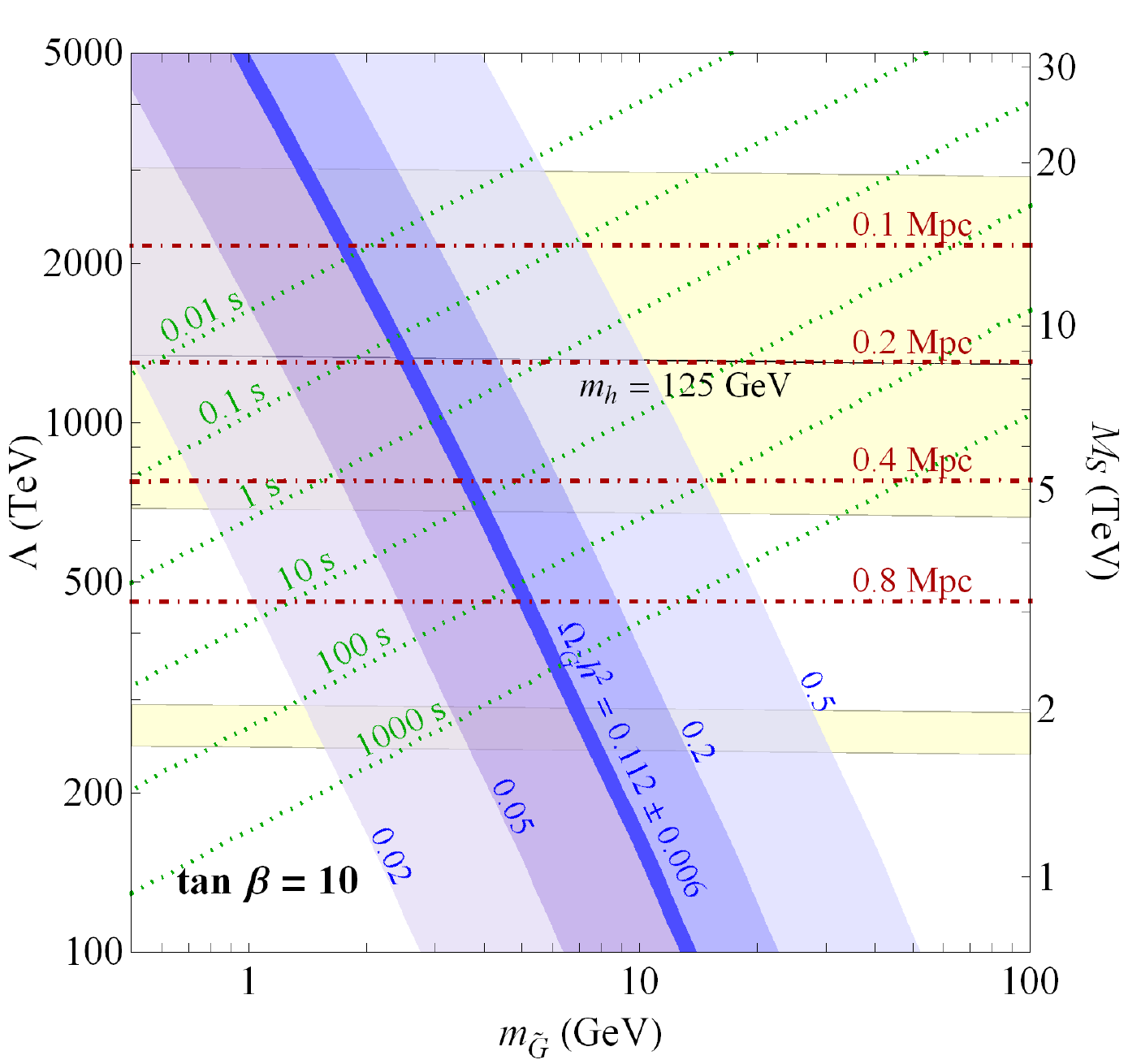}\hfil
\includegraphics[width=0.495\textwidth]{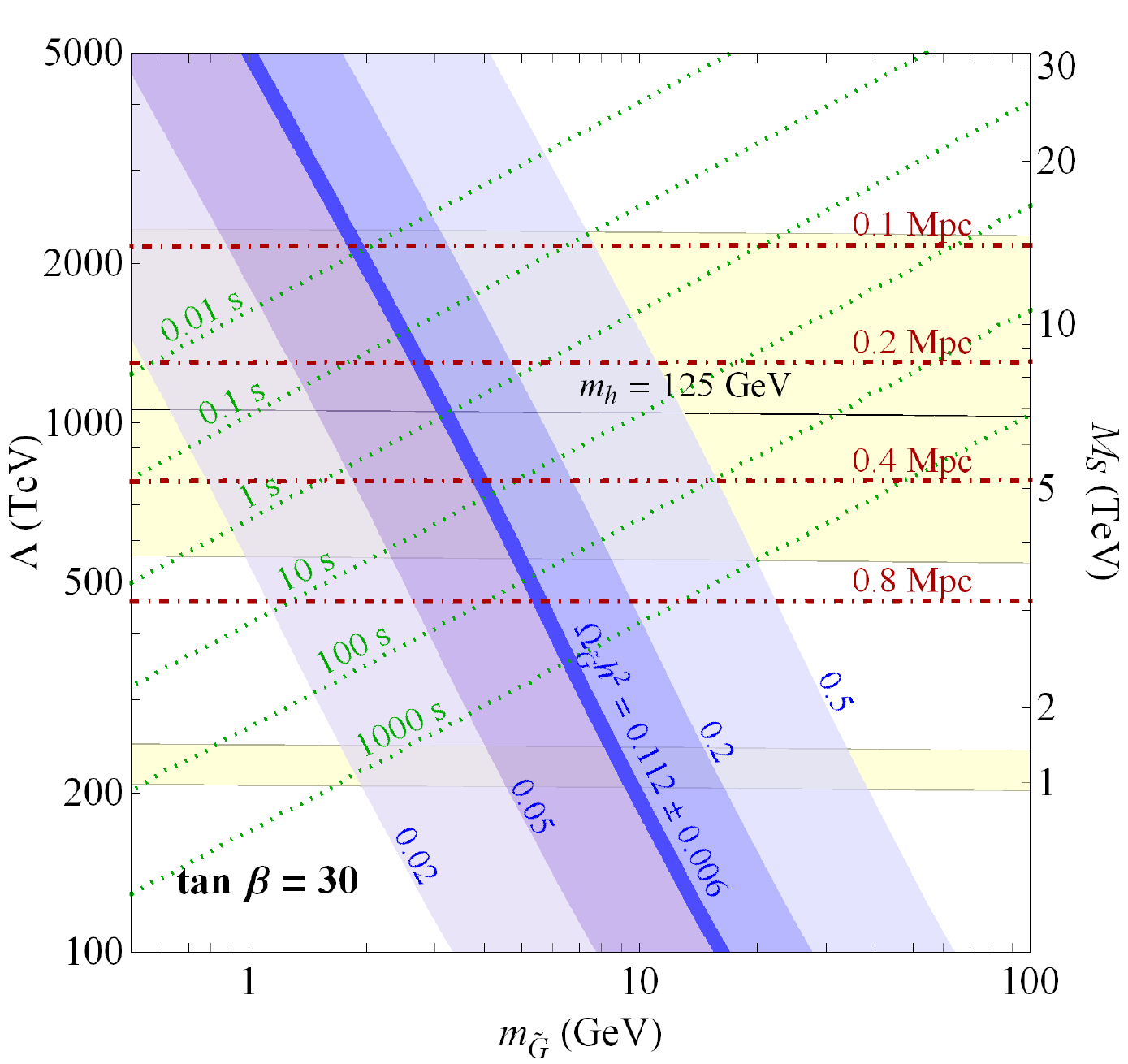}
\caption{\label{fig:relic} Contours of constant gravitino relic
  density (blue, shaded), neutralino lifetime (dotted, green) and
  free-streaming length (dot-dashed, red) for $\tan \beta = 10$ (left)
  and 30 (right), $N_5 = 1$, and $\mu > 0$.  The preferred Higgs mass
  regions from \figref{higgsedm} are also shown. The stop mass
  parameter $M_S$ is largely determined by $\Lambda$ and insensitive
  to $\mg$ in the range plotted, and it is given on the right-hand
  axis. }
\end{figure}

Now we turn to the cosmological constraints on this dark matter
scenario.  Since the gravitino couples to the neutralino through its
Goldstino component, its coupling is suppressed by $1/F$ and the
neutralino may have a long lifetime. Neglecting the mass of the $Z$
boson with respect to that of the neutralino, we estimate the neutralino's
lifetime as
\begin{equation}
   \label{lifetime}
   \tau_\chi\simeq\frac{48\pi\mg^2\mpl^2}{m^5_\chi}
   \simeq0.02~\text{sec}\left(\frac{m_{\tilde{G}}}
   {1~\gev}\right)^2\left(\frac{2~\tev}{\mx}\right)^5 \ ,
\end{equation}
where we have included both the $\gamma\tilde G$ and $Z\tilde G$ decay
channels. 

Such late production of dark matter is constrained by various
astrophysical and cosmological observations. Daughter particles from
neutralino late decays deposit energy to the plasma in the early
universe and lead to potentially observable effects.  There are many
constraints on late energy injections, such as entropy production, the
cosmic microwave background, and
BBN~\cite{Feng:2003xh,Feng:2003uy}. For the model we consider here,
the bound from BBN is the most stringent~\cite{Feng:2008zza}, and it
requires the neutralino lifetime to be less than $\sim 0.1 - 1
~\text{s}$.  In \figref{relic}, we show contours of constant
neutralino lifetime in the $(\mg, \Lambda)$ plane for two values of
$\tan\beta$.  The BBN constraints exclude regions of parameter space
with low $\Lambda$, but are consistent with the Higgs-preferred values
of $\Lambda \agt 500 - 2000~\tev$, depending on $\mg$.

Another important constraint on dark matter produced in late decays is
from considerations of small scale structure~\cite{Borgani:1996ag,%
  Lin:2000qq,Hisano:2000dz,Kaplinghat:2005sy,Cembranos:2005us,%
  Jedamzik:2005sx,Borzumati:2008zz}.  Since the gravitino is much
lighter than the neutralino in the preferred region, it is
relativistic when it produced.  Moreover, it is produced at late
times, when the Hubble expansion rate has decreased and the redshift
effect is not efficient in reducing the gravitino velocity
significantly.  Thus the late-produced gravitino may have a large
free-streaming length and hence suppress structure on small scales.
The free-streaming length $\lambda_{\text{FS}} =
\int^{t_{\text{EQ}}}_\tau dt v(t)/a(t)$ is approximated by
\begin{eqnarray}
\label{fs}
   \lambda_{\text{FS}} \simeq 1.0~\Mpc \left[\frac{u^2_\tau\tau}
   {10^6\s}\right]^{1/2}\left[1-0.07\ln\left(\frac{u^2_\tau\tau}
   {10^6~\s}\right)\right] \ ,
\end{eqnarray}
where
\begin{equation}
   u_\tau\equiv \frac{|\vec{p}_{\tilde{G}}|}{\mg} 
\approx \frac{\mx}{2\mg}
\end{equation}
is evaluated at the decay time $\tau$.  Note that the free-streaming
length is independent of $\mg$.  As evident from \eqref{fs},
$\lambda_{\text{FS}}$ depends only on $u_{\tau}^2 \tau$, but since
$u_\tau \propto 1/\mg$ and $\tau\propto\mg^2$, the dependence on the
gravitino mass cancels.  Current constraints require
$\lambda_{\text{FS}} \alt 0.5~\Mpc$, but values near this bound may,
in fact, be preferred by observations. Values for
$\lambda_{\text{FS}}$ are also shown in \figref{relic}.  Constraints
on $\lambda_{\text{FS}}$ again exclude low values of $\Lambda$, but
are consistent with the values $\Lambda \agt 1000~\tev$ required to
produce the desired Higgs boson mass.

All of the particle physics and cosmological constraints discussed so
far are summarized in \figsref{cosmoedm}{summary}. In
\figref{cosmoedm}, we simplify the presentations of bounds in previous
figures by selecting a contour for each observable that can be thought
of as the boundary between the excluded and viable regions of
parameter space. We require that the electron's EDM bound be satisfied
for $\tan \thCP = 0.1$, the correct relic density
$\Omega_{\tilde{G}}h^2 = 0.112 \pm 0.006$, the neutralino lifetime to
be $\tau < 1~\s$ to avoid ruining BBN successes, and the gravitino to
be sufficiently cold, with $\lambda_{\text{FS}} < 0.5~\Mpc$.  Finally,
we also show the regions with the Higgs mass in the currently allowed
range.  Note that uncertainties from the experimental measurement, the
theoretical calculation of $m_h$ in supersymmetry, and parametric
uncertainties from uncertainties in $\alpha_s$ and $m_t$ are all at
the few GeV level.  Within the uncertainties that enter this and the
other observables, however, it is a remarkable fact that all of the
constraints may be satisfied in the region of minimal GMSB parameter
space corresponding to a Higgs boson mass in the currently allowed
range.

\begin{figure}[tbp]
\includegraphics[width=0.495\textwidth]{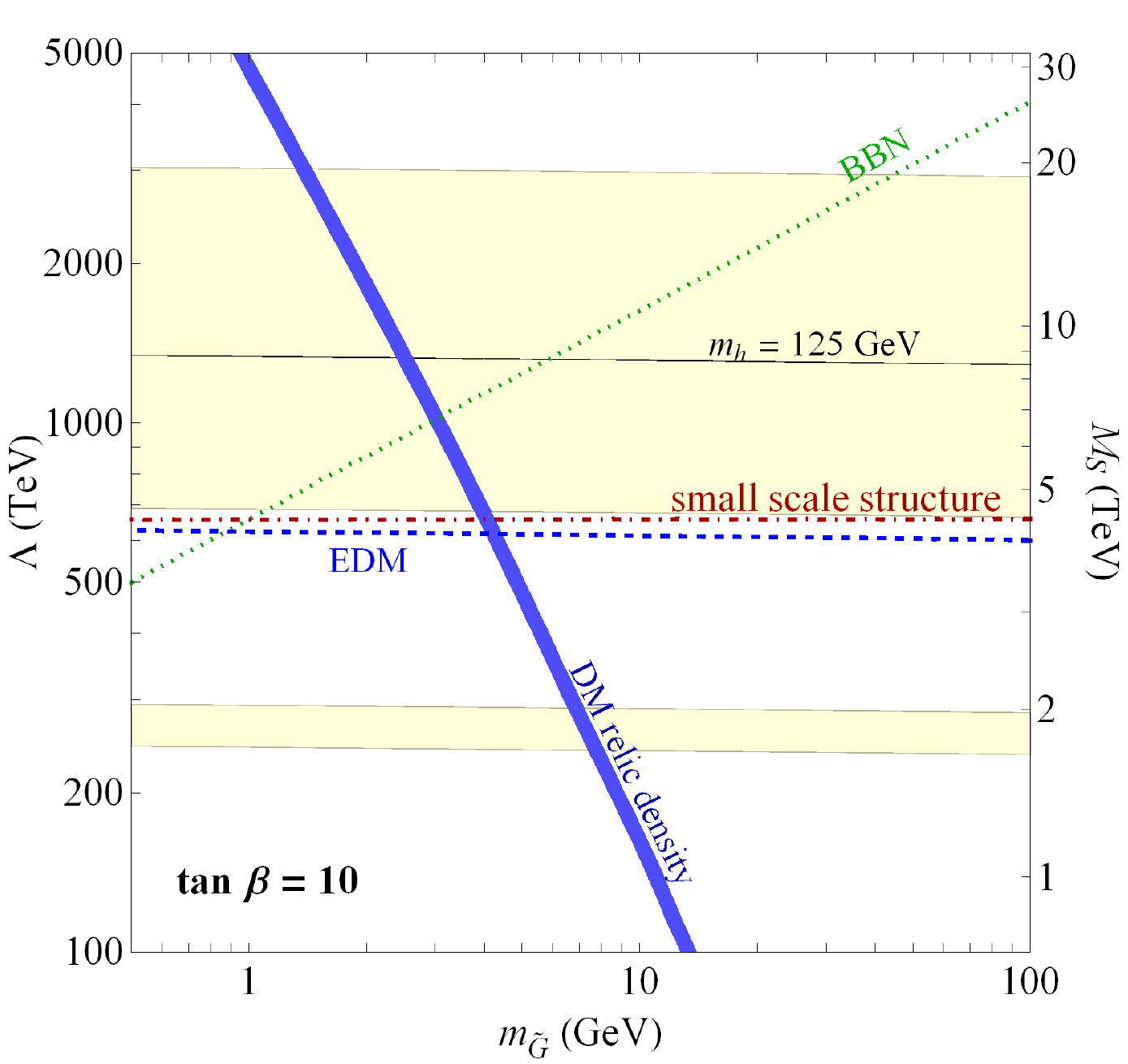}
\hfil
\includegraphics[width=0.495\textwidth]{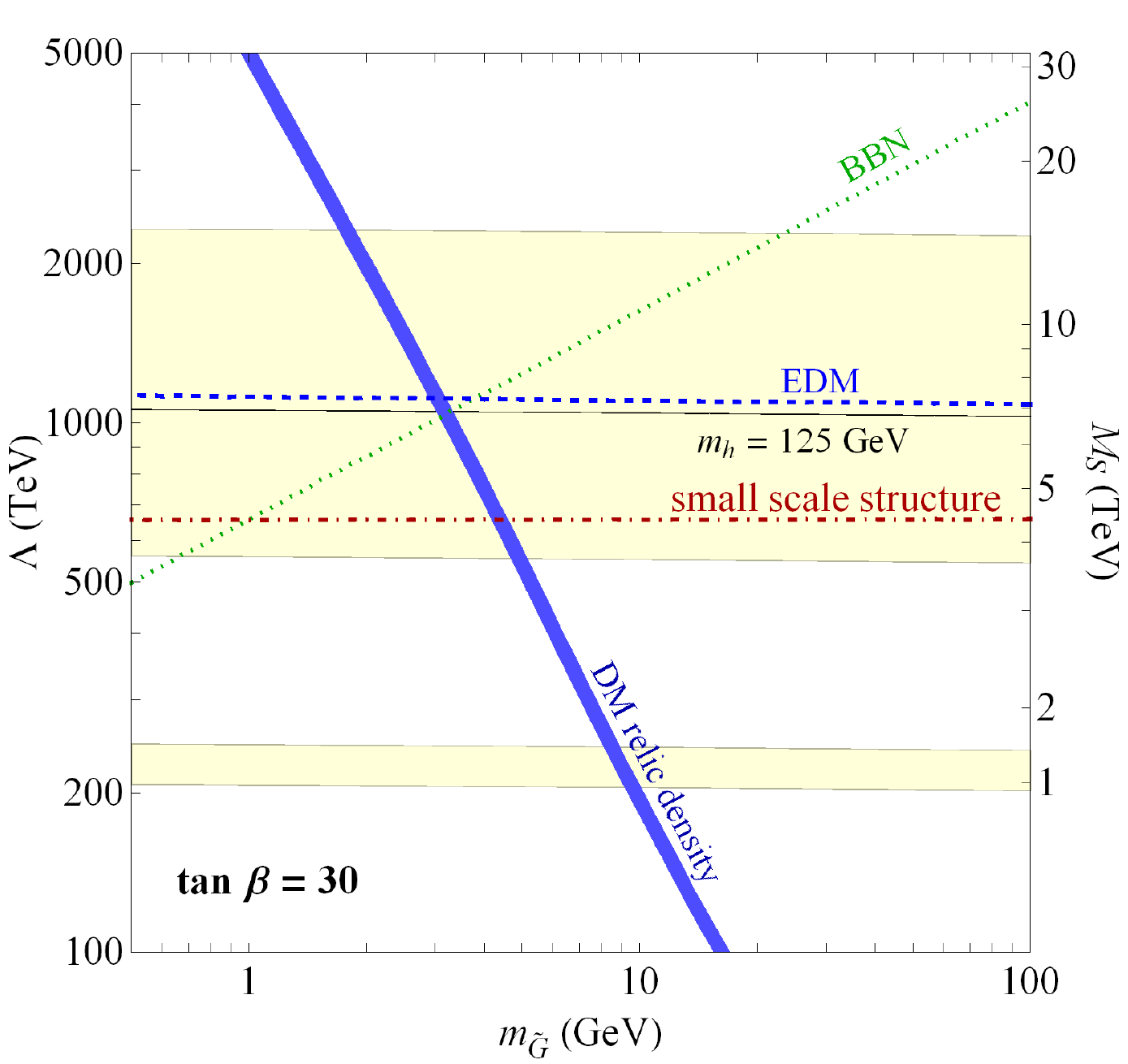}
\caption{\label{fig:cosmoedm} Summary plot of all constraints on the
  minimal GMSB scenario in the $(\mg, \Lambda)$ plane for $\tan \beta
  = 10$ (left) and 30 (right), $N_5 = 1$, and $\mu>0$.  The Higgs mass
  is in the allowed range in the light yellow shaded regions, and
  $\Omega_{\tilde{G}}h^2=0.112 \pm 0.006$ in the dark blue shaded
  bands.  The EDM constraint (for $\tan\thCP=0.1$) and BBN
  constraint ($\tau_\chi < 1~\s$) exclude parameter space below the
  indicated contours, and small-scale structure ($\lambda_{\text{FS}}
  \alt 0.5~\Mpc$) favors parameter space above or near the indicated
  contour. The stop mass parameter $M_S$ is largely determined by
  $\Lambda$ and insensitive to $\mg$ in the range plotted, and it is
  given on the right-hand axis. }
\end{figure}

\begin{figure}[tbp]
\centering
\includegraphics[width=0.495\textwidth]{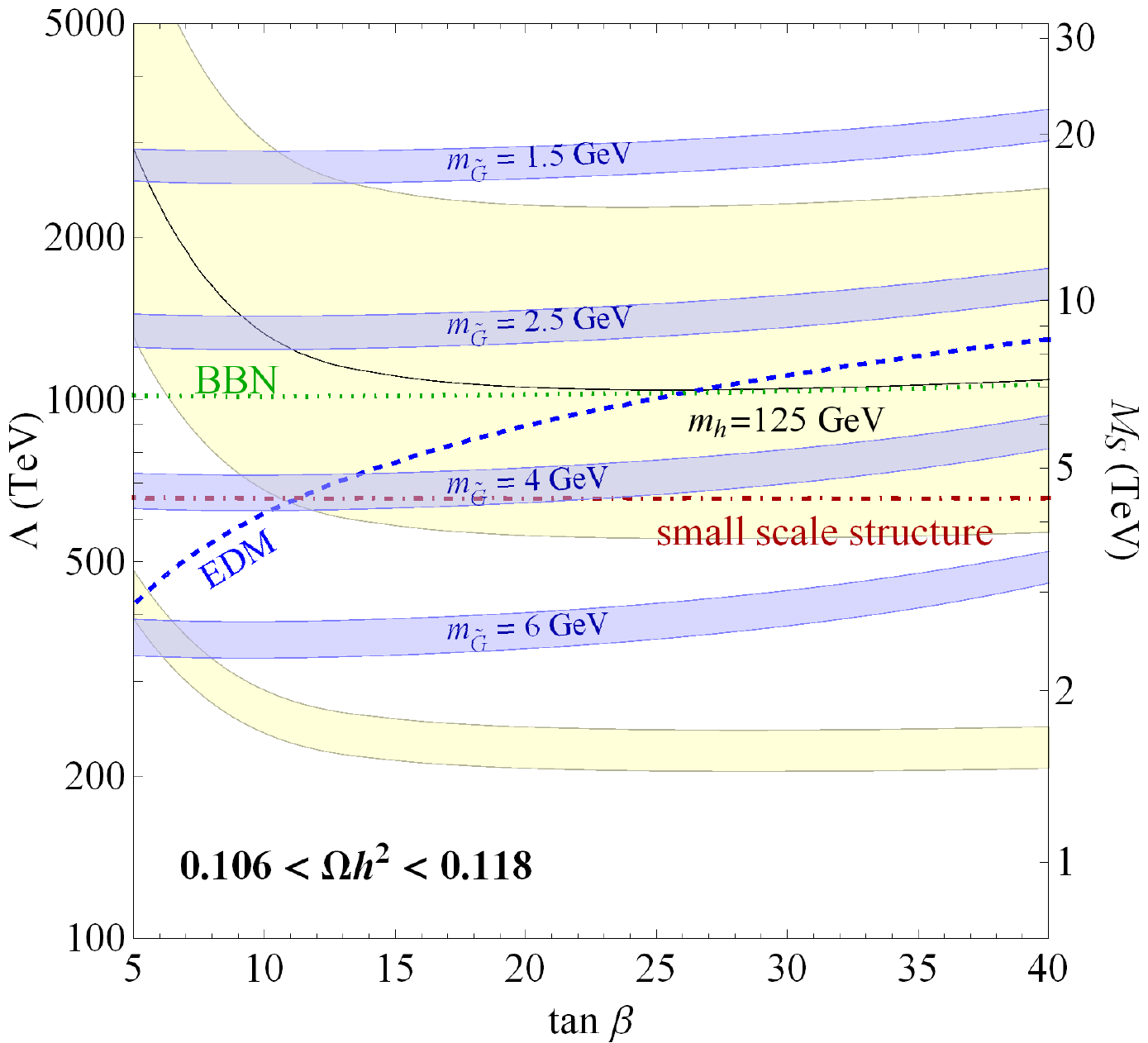}
\caption{\label{fig:summary} Summary plot of all constraints on the
  minimal GMSB model in the $(\tan \beta, \Lambda)$ plane for $N_5 =
  1$, $\mu>0$. The blue-shaded bands correspond to regions where the
  correct gravitino relic density is achieved for the indicated
  gravitino mass. Constraints from EDMs, BBN, small scale structure,
  and the Higgs boson mass are as in \figref{cosmoedm}.  The stop mass
  parameter $M_S$ is largely determined by $\Lambda$ and insensitive
  to $\tan\beta$, and it is given on the right-hand axis.}
\end{figure}

In \figref{summary}, we present an alternative summary view of our
results, in which we scan over a wide range of $\tan\beta = 5 - 40$.
For a given point in the resulting $(\tan \beta, \Lambda)$ parameter
space, $\mg$ is set by requiring the current gravity relic
density. The required $\Lambda$ for four representative values for
$\mg$ are shown.  All points in this parameter space therefore have
the correct relic density, and constraints from the various
observables are then shown, as in \figref{cosmoedm}.  Although some of
the constraints, notably those from $m_h$ and the electron EDM, have
significant dependence on $\tan\beta$, we again see that, within the
uncertainties associated with the various observables, all of the
constraints may be simultaneously satisfied in regions of parameter
space indicated by current hints for Higgs boson discovery.

\section{Summary and Discussion}
\label{sec:conc}

Recent LHC results provide tantalizing hints that a SM-like Higgs
boson exists at a mass around 125 GeV.  These results have strong
implications for supersymmetry, where they require stops with
multi-TeV masses or significant left-right mixing.  In this work, we
have considered the framework of GMSB, in which flavor violation is
elegantly suppressed.  GMSB models typically have little left-right
mixing, however, and so require multi-TeV stops to raise the Higgs
boson mass.  We have shown that such masses are highly motivated from
other perspectives.  In particular, they adequately suppress the EDMs,
even for ${\cal O}(1)$ phases, and they allow for a solution to the
dark matter problem in GMSB in the form of Goldilocks cosmology.  In
this scenario, TeV neutralinos freezeout with large densities, but
then decay to GeV gravitinos, which have the correct relic density to
be all of dark matter.  This dark matter scenario brings with it its
own set of additional constraints from the relic density, BBN, and
small scale structure.  Remarkably, we have shown that within the
uncertainties that enter these constraints, all of them, from low
energy particle physics, colliders, cosmology, and astrophysics, are
satisfied in the region of minimal GMSB parameter space corresponding
to a Higgs boson mass in the currently allowed range.

The model we have analyzed accommodates the 125 GeV Higgs boson
without additional fields and without modifications to standard big
bang cosmology.  If the hints for a SM-like 125 GeV Higgs boson are
born out, this will be among the simplest and most minimal of
supersymmetric explanations.  How can it be verified?  The resulting
spectrum has squark, gluino, and heavy Higgs masses around 5 TeV or
above, and slepton, chargino, and neutralino masses around 1 to 4 TeV.
Two example spectra are shown in \figref{spectrum}.  Such particles
are unlikely to be seen at the 14 TeV LHC, and will be extremely
challenging to discover even at future colliders.

\begin{figure}[tbp]
\centering
\includegraphics[width=0.495\textwidth]{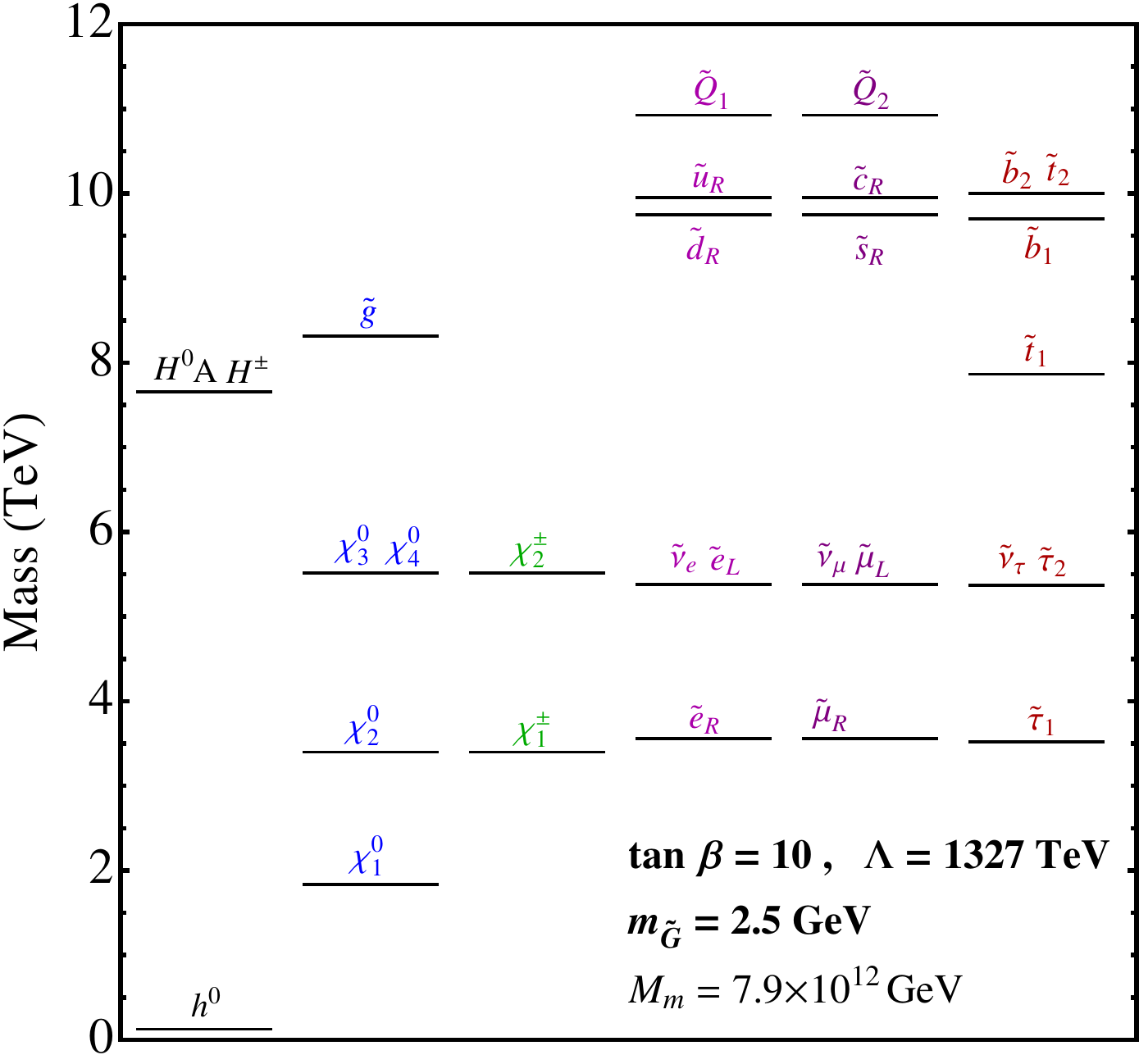}
\hfil
\includegraphics[width=0.495\textwidth]{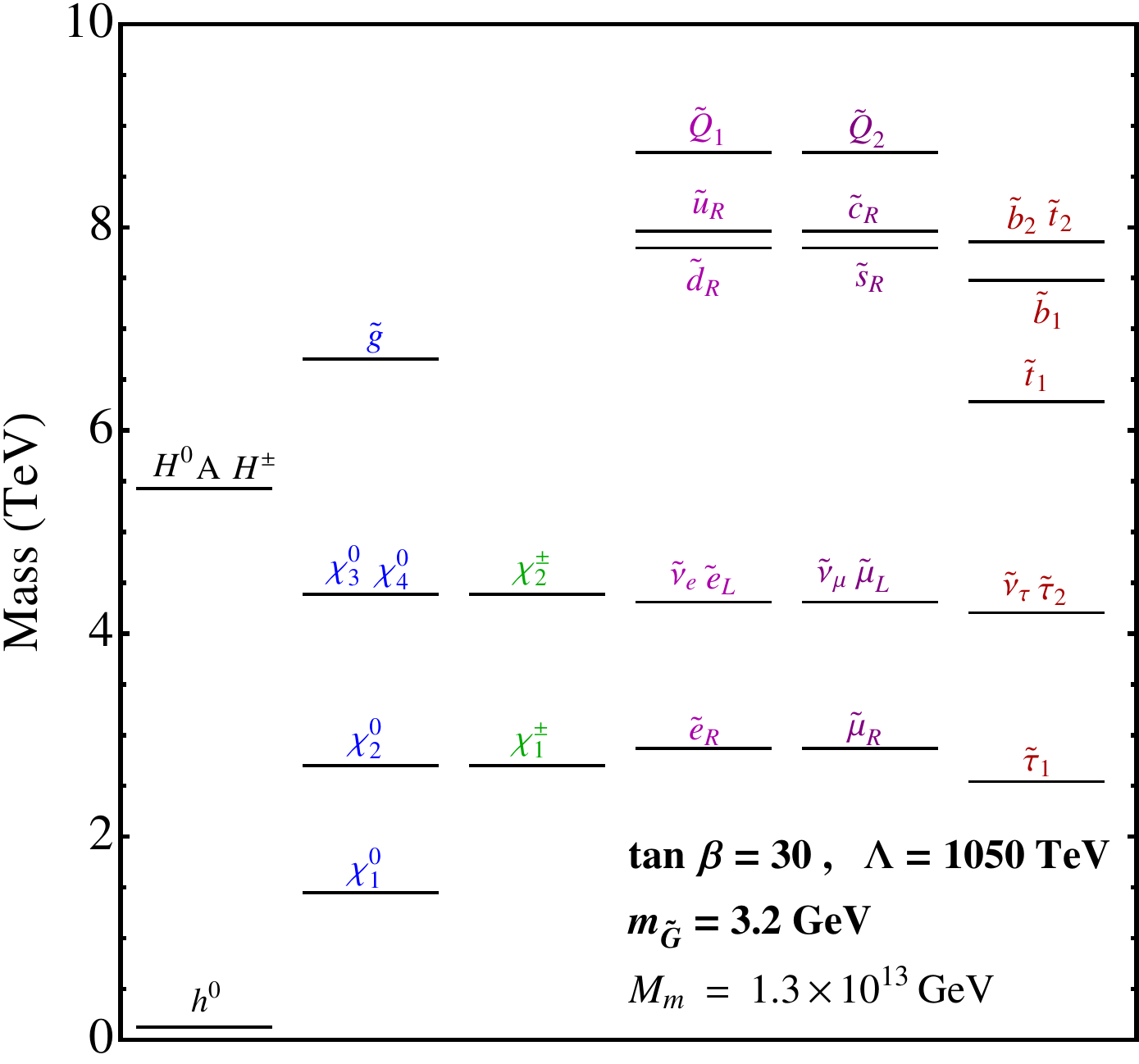}
\caption{\label{fig:spectrum} Superpartner mass spectra for example
  models with $\tan\beta = 10$ (left) and 30 (right), $\Lambda$, $m_{\tilde{G}}$
  and $M_m$ as indicated, and $N_5 = 1$ and $\mu > 0$.
  For each case, the parameters $\Lambda$ and $M_m$ have been uniquely
  fixed by requiring that $m_h = 125~\gev$ and $\Omega_{\tilde{G}} = 0.112$.
  For the $\tan\beta = 10$ example, $\tau_{\chi} = 0.18~\s$,
  $\lambda_{\text{FS}} = 0.19~\Mpc$, and $\text{max} ( \tan
  \theta_{\text{CP}}) = 0.43$. For the $\tan\beta = 30$ example,
  $\tau_{\chi} = 0.92~\s$, $\lambda_{\text{FS}} = 0.27~\Mpc$, and
  $\text{max} ( \tan \theta_{\text{CP}}) = 0.090$.  }
\end{figure}

The EDMs provide a more promising possibility.  In the region of
parameter space corresponding to a 125 GeV Higgs boson, the values of
the electron and neutron EDMs are not far from their current bounds.
There are many proposed experiments that will improve current bounds,
in some cases by 2 or 3 orders of magnitude (see Section 7.2 of
Ref.~\cite{Hewett:2012ns}).  The prediction of the models studied here
is that, assuming ${\cal O}(1)$ phases, non-zero values for the EDMs
will be discovered with the next order-of-magnitude improvement.

Cosmological studies may also shed light on this scenario.  As shown
above, the parameter space corresponding to a 125 GeV Higgs implies
free-streaming lengths in the range $\lambda_{\text{FS}} \agt
0.1~\Mpc$.  Such dark matter may explain current hints that the dark
matter is not cold, but warm.

Finally, as noted above, multi-TeV stops are typically considered
unnatural. Naturalness is, of course, quite subjective and there are
well-known mechanisms by which the sensitivity of the weak scale to
variations in the fundamental parameters can be reduced (see, for
example,
Refs.\cite{Feng:1999mn,Feng:1999zg,Agashe:1999ct,Kitano:2005wc,Feng:2012jf}.
We note, however, that, although the case of sub-TeV stops with
significant left-right stop mixing might appear more natural, as shown
in this study, even in flavor-conserving frameworks, the CP-violating
EDMs typically require multi-TeV scalars.  Models advanced to resolve
the conflict between naturalness and the Higgs constraints with
sub-TeV stops are incomplete unless they simultaneously also explain
the suppression of low-energy flavor and CP violation and the origin
of dark matter.

\begin{acknowledgments}
We thank Kevork Abazajian, James Bullock, Manoj Kaplinghat and
David Sanford for useful discussions. JLF and ZS are
supported in part by NSF grant PHY-0970173. HBY is supported in part
by NSF grant PHY-1049896 and by NASA Astrophysics Theory grant
NNX11AI17G.
\end{acknowledgments}

\vspace{0.15in}
\noindent {\em Note added}: As this work was being submitted, a
paper~\cite{Okada:2012} appeared that also discusses the implications
of a 125 GeV Higgs boson for GMSB.


\providecommand{\href}[2]{#2}\begingroup\raggedright\endgroup

\end{document}
